\documentclass[aip,rsi,reprint,graphicx]{revtex4-1} 

\usepackage[pdftex]{graphicx}
\graphicspath{{figures/}}

\usepackage{epstopdf}
\usepackage{mathtools}
\usepackage{wasysym}
\usepackage{natbib}
\usepackage{afterpage}
\usepackage[usenames,dvipsnames,svgnames,table]{xcolor}
\usepackage{amsmath}
\usepackage{color}
\usepackage{xcolor}

\renewcommand{\deg}{^\circ}

\begin{document}

\title{A facility for the analysis of the electronic structures of solids and their surfaces by synchrotron radiation photoelectron spectroscopy} 

\author{M.~Hoesch}
\email[]{Moritz.Hoesch@diamond.ac.uk}
\affiliation{Diamond Light Source, Harwell Science \& Innovation Campus, Didcot, OX11 ODE, United Kingdom}

\author{T.~K.~Kim}
\affiliation{Diamond Light Source, Harwell Science \& Innovation Campus, Didcot, OX11 ODE, United Kingdom}

\author{P.~Dudin}
\affiliation{Diamond Light Source, Harwell Science \& Innovation Campus, Didcot, OX11 ODE, United Kingdom}

\author{H.~Wang}
\affiliation{Diamond Light Source, Harwell Science \& Innovation Campus, Didcot, OX11 ODE, United Kingdom}

\author{S.~Scott}
\affiliation{Diamond Light Source, Harwell Science \& Innovation Campus, Didcot, OX11 ODE, United Kingdom}

\author{P.~Harris}
\affiliation{Diamond Light Source, Harwell Science \& Innovation Campus, Didcot, OX11 ODE, United Kingdom}

\author{S.~Patel}
\affiliation{Diamond Light Source, Harwell Science \& Innovation Campus, Didcot, OX11 ODE, United Kingdom}

\author{M.~Matthews}
\affiliation{Diamond Light Source, Harwell Science \& Innovation Campus, Didcot, OX11 ODE, United Kingdom}

\author{D.~Hawkins}
\affiliation{Diamond Light Source, Harwell Science \& Innovation Campus, Didcot, OX11 ODE, United Kingdom}

\author{S.~G.~Alcock}
\affiliation{Diamond Light Source, Harwell Science \& Innovation Campus, Didcot, OX11 ODE, United Kingdom}

\author{T.~Richter}
\affiliation{Diamond Light Source, Harwell Science \& Innovation Campus, Didcot, OX11 ODE, United Kingdom}
\affiliation{Data Management and Software Centre, European Spallation Source ERIC, Ole Maal\o{}es Vej 3, 2200 Copenhagen, Denmark}

\author{J.~J.~Mudd}
\affiliation{Diamond Light Source, Harwell Science \& Innovation Campus, Didcot, OX11 ODE, United Kingdom}

\author{M.~Basham}
\affiliation{Diamond Light Source, Harwell Science \& Innovation Campus, Didcot, OX11 ODE, United Kingdom}

\author{L.~Pratt}
\affiliation{Diamond Light Source, Harwell Science \& Innovation Campus, Didcot, OX11 ODE, United Kingdom}

\author{P.~Leicester}
\affiliation{Diamond Light Source, Harwell Science \& Innovation Campus, Didcot, OX11 ODE, United Kingdom}

\author{E.~C.~Longhi}
\affiliation{Diamond Light Source, Harwell Science \& Innovation Campus, Didcot, OX11 ODE, United Kingdom}

\author{A.~Tamai}
\affiliation{Department of Quantum Matter Physics, University of Geneva, 24 Quai Ernest-Ansermet, 1211 Geneva 4, Switzerland}

\author{F.~Baumberger}
\affiliation{Department of Quantum Matter Physics, University of Geneva, 24 Quai Ernest-Ansermet, 1211 Geneva 4, Switzerland}
\affiliation{Swiss Light Source, Paul Scherrer Institut, CH-5232 Villigen PSI, Switzerland}

\date{\today}

\begin{abstract}
A synchrotron radiation beamline in the photon energy range of  18 - 240 eV and an electron spectroscopy end station have been constructed at the 3~GeV Diamond Light Source storage ring. The instrument features a variable polarisation undulator, a high resolution monochromator, a re-focussing system to form a beam spot of 50x50~$\mu$m$^2$ and an end station for angle-resolved photoelectron spectroscopy (ARPES) including a 6-degrees-of-freedom cryogenic sample manipulator.  The beamline design and its performance allow for a highly productive and precise use of the ARPES technique at an energy resolution of 10 - 15 meV for fast $k$-space mapping studies with a photon flux up to $2\cdot 10^{13}$ ph/sec and well below 3~meV for high resolution spectra.

\end{abstract}


\maketitle

\section{Introduction}

Photoelectron spectroscopy is a highly versatile tool for the investigation of the electronic structure of solids and their surfaces. The specific regime of very high energy resolution angle-resolved photoelectron spectroscopy (ARPES) has proven to be particularly successful for the measurement of electrons at very low binding energies in crystalline correlated electron systems~\cite{hufnerbook, damascellireview, campuzanoreview} and is thus complementary to transport measurements on one hand and to (scanning) tunnelling spectroscopy on the other. Diamond Light Source has built a VUV to soft x-ray beamline, which, combined with an end station for high energy and angular resolution photoelectron spectroscopy forms a highly productive facility.  Measurements are performed on both, cleaved single crystals with high through-put, as well as samples grown and prepared by complex surface science and molecular beam epitaxy methods in the beamline vacuum system. This instrument is unique in the coherent approach taken to optimise all components into a well-balanced ensemble. The design choices and design considerations for the beamline have been described in Ref. [\onlinecite{hoesch16}].

\begin{figure*}[!htbp]
	\includegraphics[width = .93\textwidth]{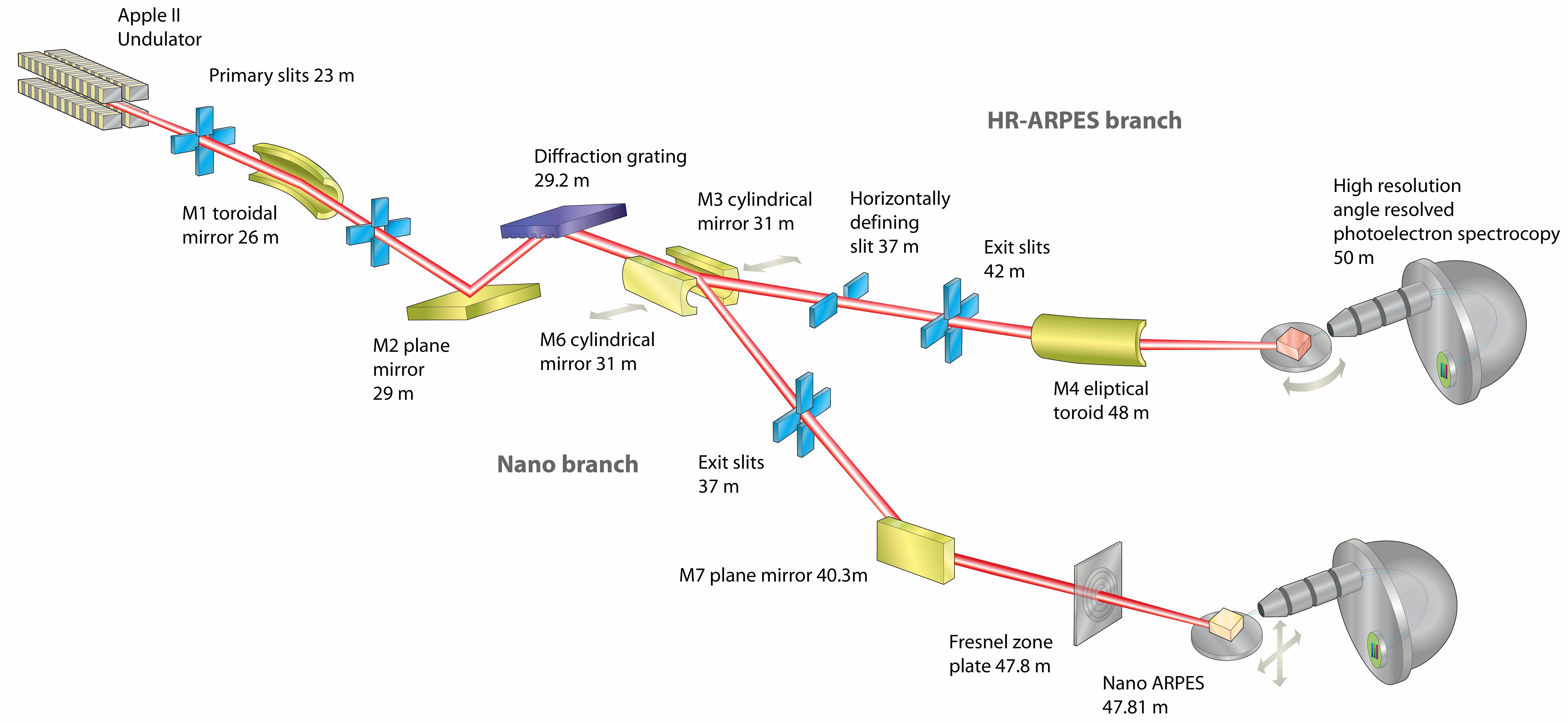}
	\caption{Schematic layout of the optics at beamline I05-ARPES consisting of the undulator source, the common branch up to the plane grating monochromator and the two branches HR-ARPES and nano-ARPES.}
\label{Fig_layout}
\end{figure*}

The technique of ARPES at high energy resolution requires a highly monochromatic intense photon beam of photon energy $E$ and energy spread $\Delta E$, combined with an electron spectrometer, which measures photoelectrons with an energy resolution $\Delta E_{ana}$. The total energy resolution may further be influenced by other effects such as sample grounding noise ($\Delta E_{etc}$) to give a combined energy resolution of 
\begin{equation}
\Delta E_{comb} = \sqrt{(\Delta E)^2+(\Delta E_{ana})^2+(\Delta E_{etc})^2} .
\label{eq0}
\end{equation}
Since the resolving power $E/\Delta E$ of a VUV monochromator is limited, typically to values of about 20,000 (the extreme reported is 100,000)~\cite{weiss2001}, the technique uses low photon energies $E$ in the VUV range for high combined energy resolution. The beamline photon energy range was thus selected to start from $h\nu = 18$~eV. The core operation range, where the energy resolution $\Delta E_{comb}$ can be kept well below 10~meV then reaches up to approx. 100~eV, above which a slightly reduced energy resolution is used for band mapping, up to 240~eV. In addition the design allows for a beam of reduced flux at a photon energy of 500~eV, which is applied to core level studies. At all photon energies the beam can be delivered in linear horizontal or vertical (LH and LV) as well as circular left and right (CL and CR) polarisations. This paper describes the design and performance of the beamline and its high resolution end station HR-ARPES. The paper is organised in sections describing the lay-out, the undulator photon source, the beam delivery design and performance as well as the end station design and performance before concluding. Not described in this paper are the second branch for spatially resolved ARPES and its end station nano-ARPES.

\section{Lay-out and beamline infrastructure}

The beamline starts inside the concrete shielding wall with the undulator source and the  front end. Optical elements are located in a lead shielded hutch and a temperature stabilised optics cabin. The hutch, optics cabin and two end station rooms, a control room and a sample preparation room are rooms built into the Diamond experimental hall. The temperature in the optics rooms is kept stable to within $0.2 \deg$C over a typical week. Two control and instrumentation areas (CIA) adjacent to the beamline rooms provide space for electronics racks for vacuum and motion control, computers and network routers as well as pre-vacuum pumps and a system of Helium compressors for closed cycle cryostat operation.

The schematic optics lay-out of the two branches is shown in Fig.~\ref{Fig_layout}. The beam is admitted into the optics through primary slits that control the illumination of the first toroidal mirror M1. A further set of slits just after M1 acts as the beamline aperture and removes unwanted radiation from the edges of mirror M1. The beam then passes over the plane mirror and grating of the plane grating monchromator (PGM) and over the cylindrical focussing mirror M3 into the exit slit ES. Alternatively the beam could be directed into the nano-ARPES branch by inserting mirror M6 instead of M3, which can be performed by a motorised motion. The dispersive direction of the monochromator is vertical. The beam is refocussed on the sample contained in the end station vacuum vessel by a elliptical toroid mirror M4. The total length of the beamline from source to end station is 50~m, about half of which is contained in the storage ring shielding tunnel.

\section{Undulator source}

The beamline is served from the long straight section I05 of the Diamond storage ring. As photon source a 5~m long variable polarisation undulator of Apple-II type~\cite{SASAKI94} was selected. The period length is $\lambda_u = 140$~mm and the total length is 5~m, thus accommodating $n_u = 34$ periods and two half-periods at the entrance and exit. It can produce all four polarisations, LH, LV, CL and CR in the range from 18~eV upwards. Higher hamonic suppression is optimised in a quasi-periodic scheme~\cite{sasaki95} similar to the design described in Ref.~[\onlinecite{panacchione09}].

\begin{figure}[!htbp]
	\includegraphics[width = .45\textwidth]{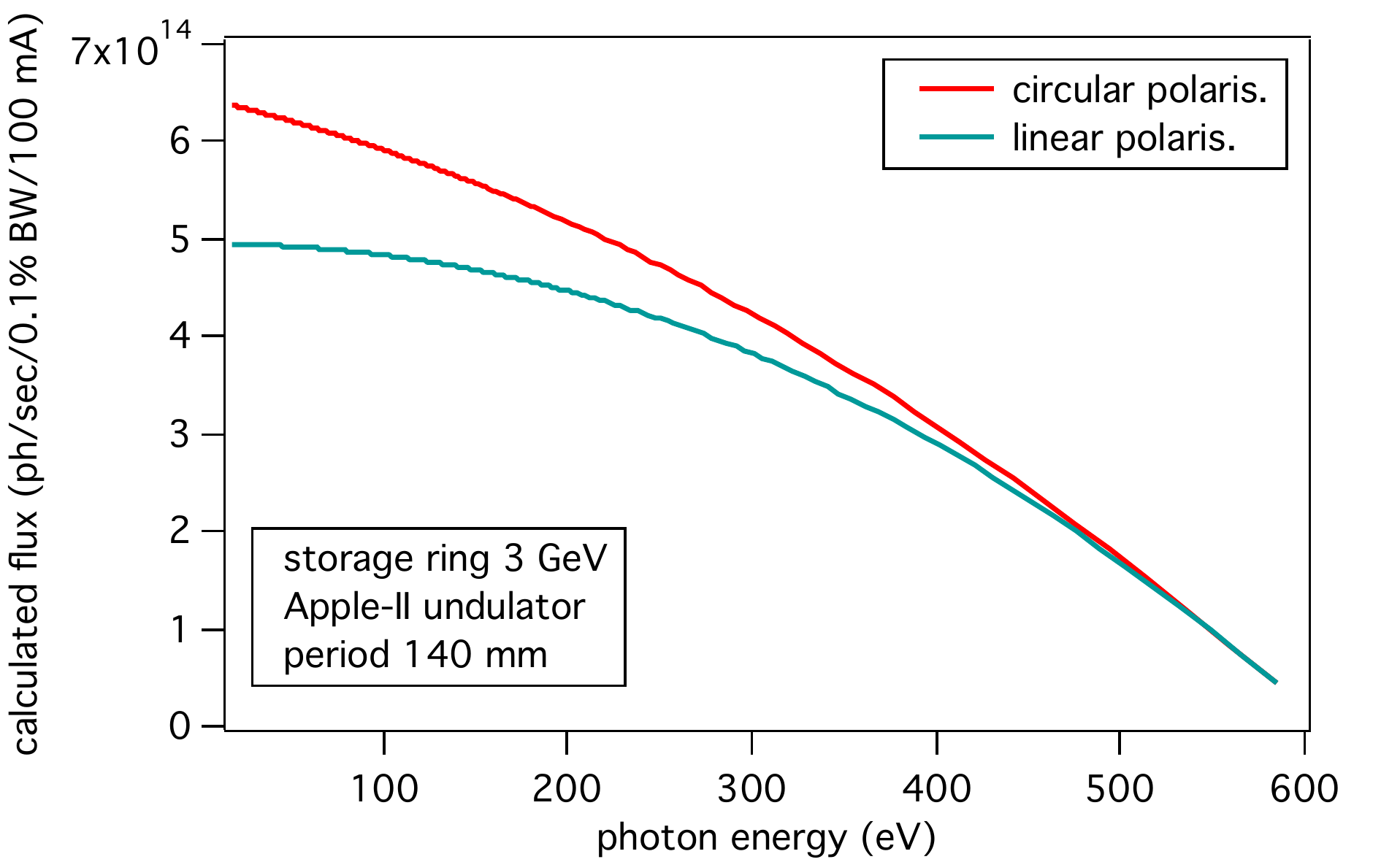}
	\caption{Calculated flux of the first harmonic of a 5m long HU140 undulator at its peak brilliance in the Diamond storage ring.}
\label{Fig_undulator_flux}
\end{figure}

\begin{figure*}[!hbtp]
	\includegraphics[width = .85\textwidth]{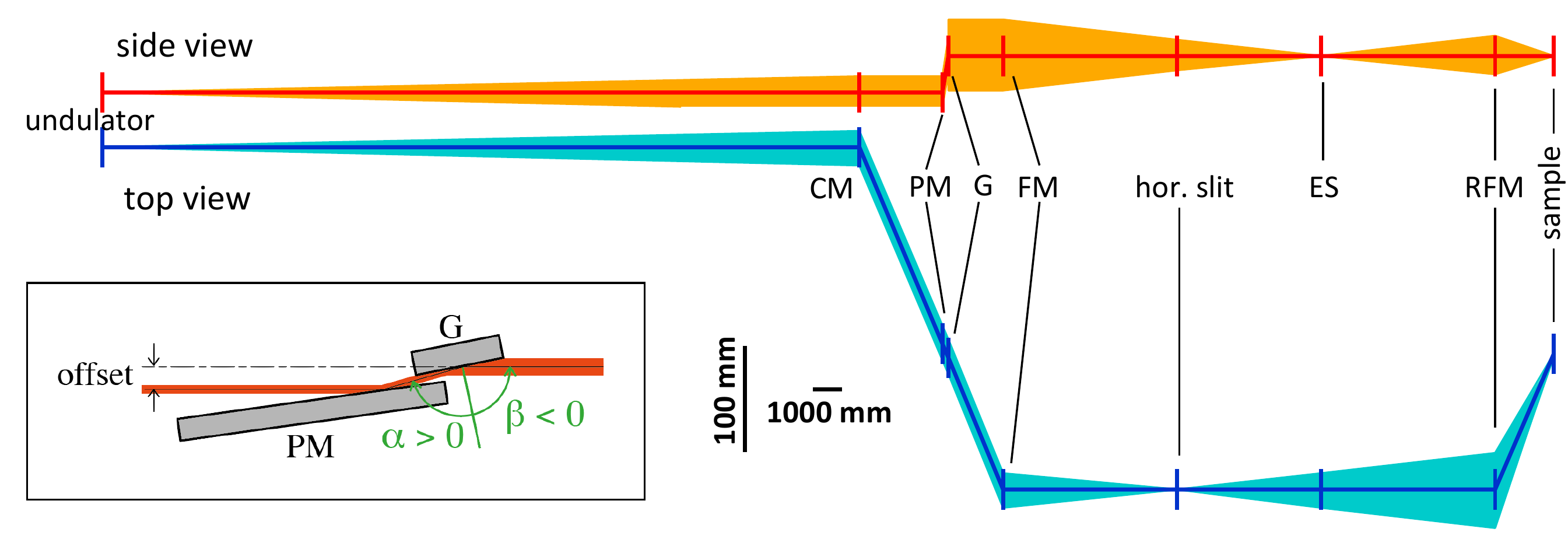}
	\caption{Scaled diagram of the optical path and beam envelope for the HR-branch of the beamline. The insert shows the mirror and grating of the PGM in 1:1 scaling.}
\label{Fig_cPGM_schematic}
\end{figure*}

To estimate the photon flux entering the beamline, Fig.~\ref{Fig_undulator_flux} shows a calculation for the fundamental of a simpler fully periodic 5m long undulator HU140. In this case the flux for LH and LV is identical. The flux is calculated at the highest brilliance point, where almost all of the beam is admitted into the beamline aperture. The intensity of the fundamental of the real undulator is slightly lower, up to 20 \%, due to the quasi-periodic perturbation,  which is not included in the prediction calculation. Note that the flux evolves very smoothly over the full range, which extends beyond 500 eV, while the beamline optics calculations below cover only the range of $18 - 240$~eV.

\section{Monochromator and refocussing optics design}

The monochromator is of collimated Plane Grating Monochromator (cPGM) type~\cite{petersen95,weiss2001,follath97,sawhney97}. This design allows a free choice of the included angle $2\vartheta = \alpha -\beta$ of the plane grating and the associated anamorphic demagnification factor $c = \cos \beta / \cos \alpha$ (see inset of Fig.~\ref{Fig_cPGM_schematic} for a definition of $\alpha$ and $\beta$). The optical functions of the beamline mirrors and slits are illustrated in Fig.~\ref{Fig_cPGM_schematic}. Collimation in the vertical plane is performed by the first toroidal mirror (CM), which also forms the horizontal intermediate focus. The focussing mirror (FM) captures the collimated beam and focusses it onto ES, at 11~m focal length, thus allowing  a distance of more than 12~m for the dispersion from the grating to the exit slit. The intermediate horizontal focus is formed by horizontal focussing of CM at 5~m upstream of ES, thus making the intermediate focus astigmatic. The refocussing mirror (RFM) is an ellliptical torus  6~m downstream of ES and 2~m before the final focus position it demagnifies vertically by 3 and horizontally by 11/2 (geometrical demagnification factors). By design the nominal horizontal beam spot size is approx.~50~$\mu$m over the whole photon energy range and with exit slit openings between 20 and 200~$\mu$m the vertical spot size varies between 7 and 70~$\mu$m. The nano-ARPES branch employs a separate FM with shorter focal length of 6~m to make a stigmatic intermediate focus  with the horizontal beam waist on the ES of this branch. 

The design uses single crystal silicon as mirror block material and the heat load is managed by water cooling, thus keeping all elements at room temperature or slightly above. At full opening of the primary slits, highest ring current and lowest gap the undulator can admit up to 1~kW of power onto the first mirror M1.  The bulk of this heat load, rather hard x-rays up to a few keV photon energy, is absorbed in M1. A deflection angle of $6 \deg$ was chosen off this mirror, which admits less than 100~W of power onto the plane mirror M2 in the PGM. Since M1 deflects horizontally, the tangential profile of the heat bump and heat deformation only affects the horizontal focussing properties. This deformation was optimised by a cross-sectional design and attachment of the cooling brackets that allows the back part of the mirror to get slightly warmer than the cooling water, thus reducing the bending of the mirror block due to temperature differences of the front surface and the bulk. M2, which has a varying incidence position along its length according to the varying angle geometry~\cite{petersen95}, can develop a heat bump, the tangential profile of which directly affects the vertical focussing plane and thus the energy resolution. This mirror employs internal water cooling by channels that run along the long length of the mirror inside the silicon and thus the heat is removed very effectively and the resulting heat bumps minimised. The internal cooling has the additional advantage that no external cooling brackets deform this 450~mm long mirror. Fully clamped into its opto-mechanical holder, using metrology feedback from the Diamond-NOM slope profilometer~\cite{alcock2010}, the mirror has a radius $> 240$~km and deviations of ~0.16~$\mu$rad (RMS) along the tangential direction. The last optical element where heat load is of concern are the gratings, where even with cooling brackets attached the slope errors are kept below 0.25~$\mu$rad~\footnote{The slope errors of gratings were measured on an equally highly polished grating blank without grooves and the procedure was optimised until the clamping of the cooling brackets was reproducible. Final slope deviations of a grating were measured in-operando with beam by partial illumination of the grating surface.}. The heat load here is small and cooling is performed by precisely aligned and lightly attached copper side-brackets with tubing that is optimised for minimum vibrations due to turbulences in the water flow. Finally the horizontally delfecting focussing mirror M3 also features a water cooled pad on the side of the mirror, which is used to reduce potential heat drift issues for the 17~m long optical arm that follows this mirror up to the reflection from M4.

\begin{figure}[!t]
	\includegraphics[width = .48\textwidth]{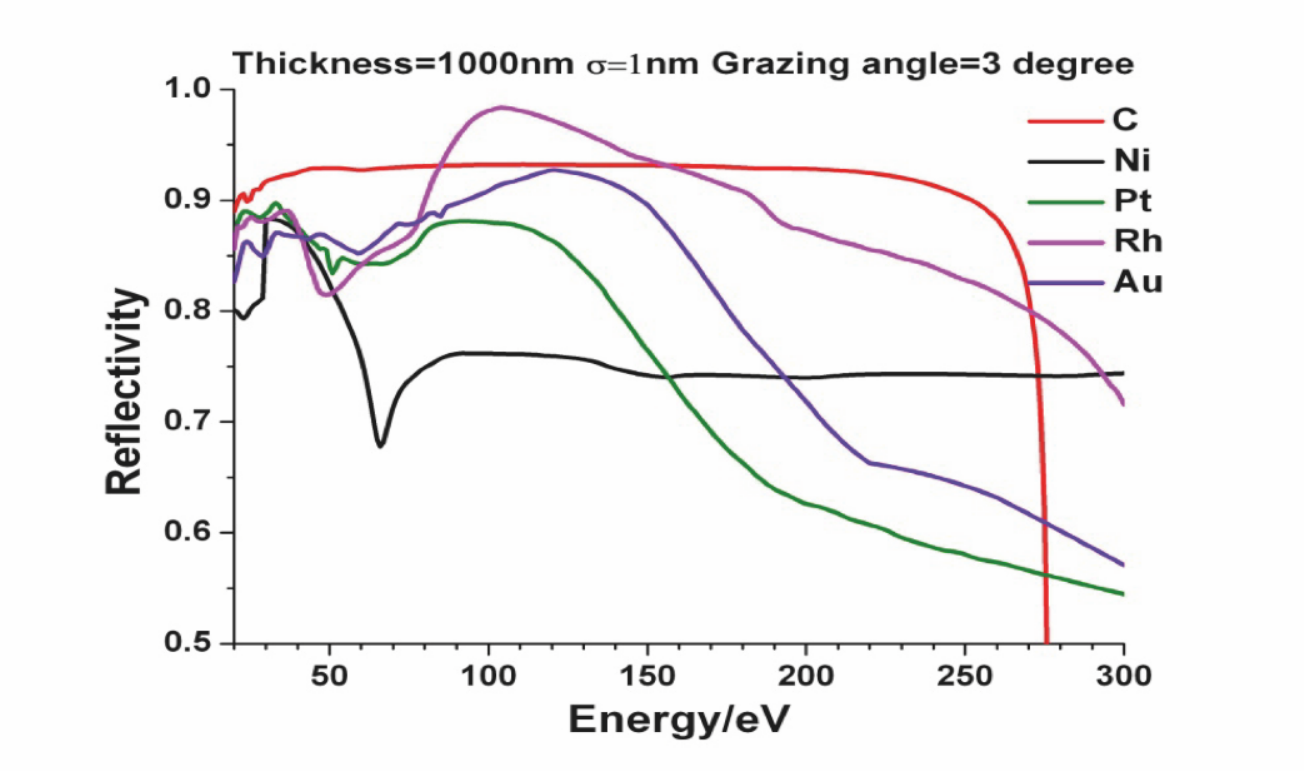}
	\caption{Calculated reflectivity for $s$-polarised radiation at an incidence angle of $3\deg$ for various mirror coating materials.}
\label{Fig_reflectivities}
\end{figure}

Mirror deflection angles were chosen as $6\deg$ for M1, M3 and M4 following a combined optimisation of the available floor space in conjunction with the choice of mirror coatings. Fig.~\ref{Fig_reflectivities} shows the calculated reflectivities of various coatings. While the highest reflectivity is found around 100~eV for a Rh coating, the amorphous carbon, calculated at a density of 2.1 g/cm$^3$, has the highest overall reflectivity and is free of energy dependent features apart from the strong drop of reflectivity on approaching the K-edge absorption (284~eV). All elements are coated with amorphous carbon, except for the gratings, where a metallic  Pt coating was chosen. M2  deflects by a variable angle between $3\deg$, used at the highest photon energies, and $36\deg$, used at the lowest photon energy where still a reflectivity of more than 60\% is expected for an $s$-polarised reflection~[\onlinecite[Sect. 4.2]{XDB2009}]. The length of the grating blocks and the plane mirror have been optimised together with the vertical beam offset in the PGM. The latter is 35~mm and with a grating block length of 150~mm and a ruled surface length of 140~mm the PGM can achieve high transmission without significant geometrical losses for an energy range of $18-600$~eV using gratings between 400 and 1200~lines/mm and $c$-values between 1.5 and 4, except for extreme combination of these.

For  operation at high resolving power $E/\Delta E > 20,000$ a grating of 800 lines/mm is used. A second grating of 400 lines/mm delivering higher transmission at still high energy resolution is also available. The optical design allows this  $E/\Delta E$ over the whole core operation range up to 100~eV above which the resolving power is slightly reduced. 
The key limiting factors of the resolution are identified as follows: (a) Residual slope errors of the plane grating and plane mirror, including cooling water-flow induced vibrations.  (b) Residual focussing errors, including slight coma aberrations of the toroidal and cylindrical mirrors. (c) Finite source size. The source size as well as additional distortions due to heat load will have a significant effect at low photon energies below $E=60$~eV, while slope errors and vibrations dominate at higher energies.

\begin{figure}[!t]
	\includegraphics[width = .49\textwidth]{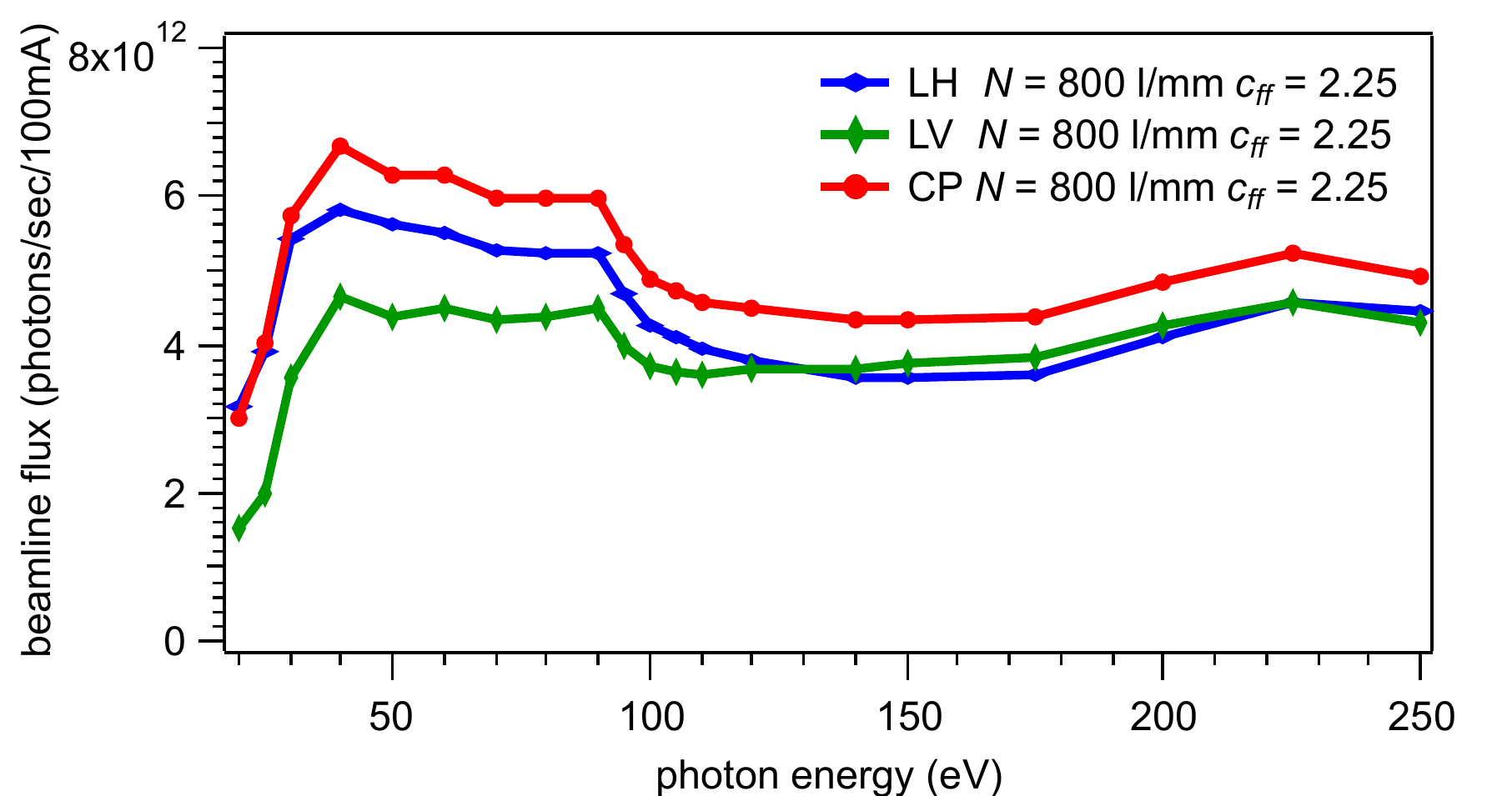}
	\caption{Estimated flux at sample on the HR-branch, calculated for an exit slit opening of 100~$\mu$m and using the source flux from Fig.~\ref{Fig_undulator_flux}.}
\label{Fig_transmission}
\end{figure}

Grating transmissions were optimised by calculations of diffraction efficiency and the groove depth as well as aspect ratio of the groove lamella was optimised. The total  beamline transmission was estimated by calculations of mirror reflectivities and grating diffraction efficiency and the resulting estimated beamline flux is shown in Fig.~\ref{Fig_transmission}. Note that the LH and LV polarised flux is slightly different due to the difference in reflectivity for $s$- and $p$-polarised reflections. This flux was calculated for an exit slit opening of 100~$\mu$m, thus corresponding to a resolving power of approx. $E/\Delta E = 6,000$. As demonstrated later, the flux is proportional to the exit slit opening due to the homogeneous illumination of ES by the dispersion of the grating.

The beamline alignment is facilitated and monitored by a set of diagnostics. Before the high heat load slits and M1 a DiagOn unit is installed with a $90\deg$ deflecting multilayer mirror optimised for an photon energy of 60~eV~\cite{diagon}. This retractable unit selectively visualises the VUV beam against the background of higher energy photons. In-situ insertable and retractable powder coated fluorescence screens are used after each horizontally confining slit unit, namely before entering the PGM, before M3 and after the horizontal intermediate focus. A further single crystal fluorescence screen is installed above the main beam axis just in front of ES, thus permanently capturing the dispersed photons and monitoring the horizontal beam position. The latter is also determined from slit blade drain current differences on the set of horizontally confining slits in the ES box, though these slits are routinely kept wide open to avoid any beam blocking in case of position drift. Further drain current measurements are provided on mirrors M3 and M4 and on the gas cell wire installed $\sim1$~m downstream of the exit slit. Photodiodes are installed on the retractable unit in the gas cell chamber as well as on a retractable diagnostics unit downstream of M4.

\section{Beam Delivery Performance}

\begin{figure*}[!thbp]
     \includegraphics[width = .95\textwidth]{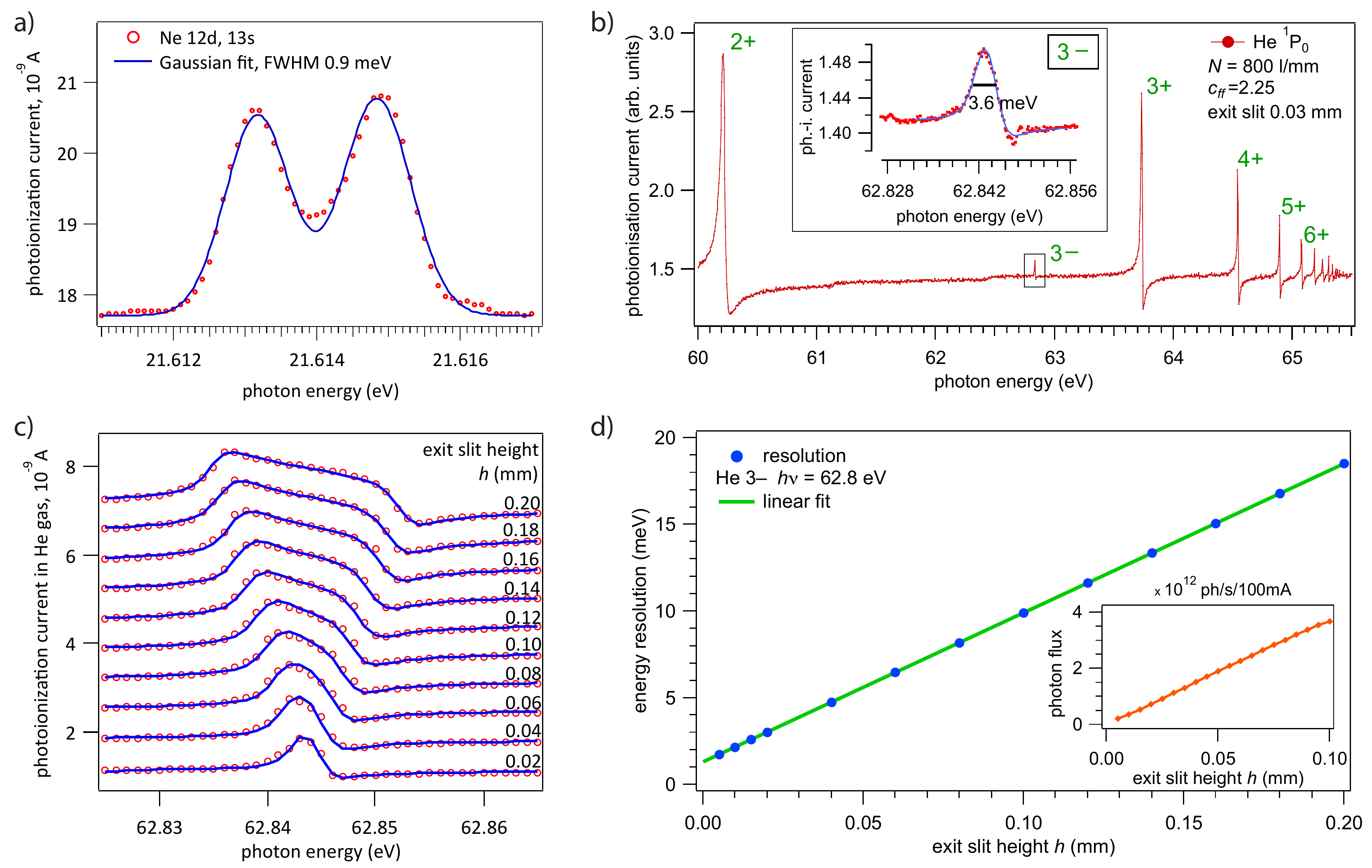}
	\caption{Measured photoionization yield spectra of (a) Ne absorption and (b) He absorption in a gas cell at an exit slit height $h=0.03$~mm using the grating of 800 lines/mm and a monochromator $c$-value of 2.25. In (b) the peaks are maked following notations in Ref.~[\onlinecite{domke96}]. (c) He gas absorption spectra at various $h$. The fits in (a), the insert of (b) and (c) are used to determine the energy resolution (fits shown as blue lines). (d) Extracted energy resolution, fit by $\Delta E = (1.3 + h\cdot 86)$~meV, with $h$ in mm. The inset shows the flux. All data are measured using LH polarisation.}
\label{Fig_energy_resol}
\end{figure*}

The beamline is aligned and focussed by a procedure similar to the one described in Ref.~[\onlinecite{strocov10}].
The beamline resolution is regularly verified at two photon energies by measuring the Fano-type resonances in photoionization spectra of He near 60 eV~\cite{domke96} and Ne gas near 20 eV~\cite{klar92,nahon01}. The pressure of the gas is set to approx. $5\cdot 10^{-2}$~mbar as measured on a Pirani vacuum gauge, and the photocurrent is picked up by a blank copper wire running parallel to the beam. The gas cell volume is isolated from the UHV of the beamline by a 150~nm thin Al membrane, that is transparent for photon energies between 20 and 70 eV~\cite{henke93}. In the case of He spectra the $2,1_{3}$ resonance from the $^1\mathrm{P}_0$ series is used due to the extremely small natural linewidth about 0.1 meV~\cite{schulz96}. In case of Ne we chose lines $12\mathrm{s} - 14 \mathrm{d}$, where each has an internal width below 10~$\mu$eV~[\onlinecite{klar92}]. Therefore the observed linewidths are always dominated by the the instrument contribution. Typical spectra with their width analysed by convolution of Fano profiles and Gaussian peaks are shown in Fig.~\ref{Fig_energy_resol}a) and b).

The required energy resolution for a given data set is adjusted by the height $h$ of ES. Fig.~\ref{Fig_energy_resol}c) shows such a typical series of He spectra. The analysis of this data set is performed by a  box function with rounded edges convoluted with the Fano function representing the natural line shape (see appendix for definition). This function is fitted to the data with the results shown in Fig.~\ref{Fig_energy_resol}d). The energy resolution falls onto a finite value of less than 2~meV at the smallest slits, thus corresponding to a $E/\Delta E > 30,000$. The smallest recommendable exit slit setting below which no substantial gain in energy resolution is achieved is  approximately $h=0.015$~mm. At these settings the beamline delivers a resolving power $E/\Delta E = 25,000$ with a flux of up to $10^{12}$~ph/sec (see Fig.~\ref{Fig_d9_test} below).

\begin{figure}[!htbp]
	\includegraphics[width = .49\textwidth]{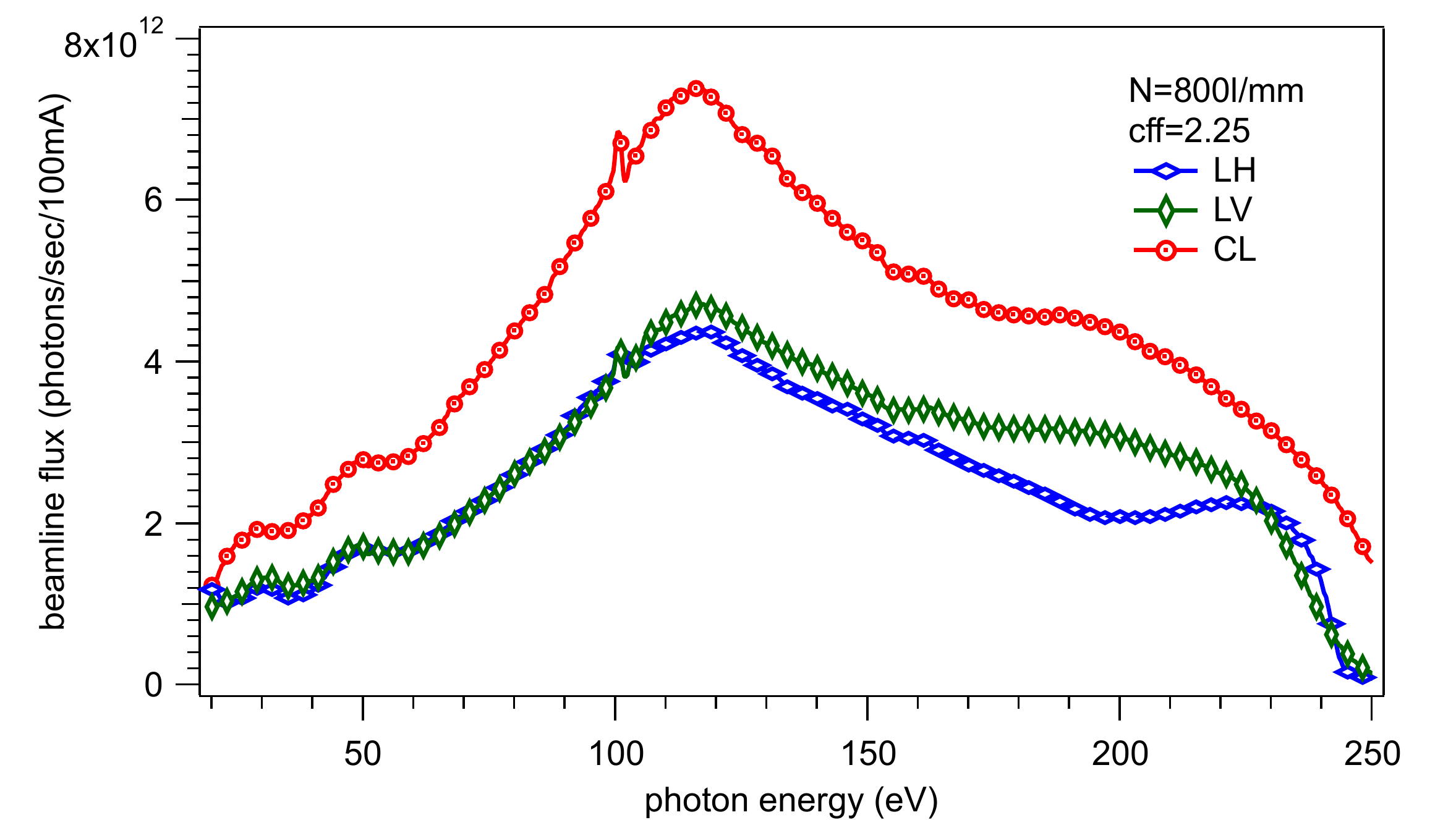}
	\caption{Measured photon flux after the last mirror of the HR-branch. The undulator and cPGM with the grating of 800 lines/mm were operated at standard settings with a 6x6 mm beamline aperture and a $c$-value of 2.25. The exit slit was set to 100~$\mu m$.}
\label{Fig_d9_test}
\end{figure}

The linear regression of the energy resolution in Fig.~\ref{Fig_energy_resol}d) can be used to eliminate the contribution of the grating dispersion to the bandwidth, and estimate the limit of instrument resolution. This yields a residual Gaussian contribution of 1.3~meV at 60~eV and 0.8~meV at 20~eV (not shown). 

The photon flux on the samples is measured by a photodiode of type SXUV100 downstream of mirror M4. The diode response was calibrated by a reference measurement at the Physikalisch Technische Bundesanstalt (PTB) in Berlin, Germany. The resulting calibrated photon flux for three different polarisation settings is shown in Fig.~\ref{Fig_d9_test}. This flux measurement may thus be affected by higher harmonic contributions and as higher photon energies contribute more to the diode current this can lead to an overestimation of the real photon flux. Furthermore the slight spike at 100~eV photon energy, corresponding to the L-edge of silicon is due to the discontinuity in the diode response combined with a slight difference in photon energy calibration between the monochromator used by PTB and the beamline. 

Overall the observed photon flux agrees well with the numbers estimated in Fig.~\ref{Fig_transmission}. The discrepancy  between LH and LV at low photon energies is less strong in reality than in the calculation. This may be due to a misestimate of the source flux, which is affected by the quasiperiodic scheme of the undulator and which may lead to higher source flux in LV than LH, while the estimate assumes identical flux with any difference coming from the difference of reflectivity.  Both CL and CR polarisations result in identical flux values within the error bars, which demonstrates excellent alignment of the undulator in the storage ring.

The photon flux curve first rises gently from low photon energies with slight humps at 30 and 50~eV and then reaches a maximum at 118~eV before it falls with a rapid drop off at 240~eV. The initial rise is due to the increased reflectivity and diffraction efficiency of the plane mirror M2 and grating at decreasing angles. The fall at high photon energies is due to the falling source flux as well as the cut-off due to the absorptions associated with the K-edge of carbon, the mirror coating (284~eV). The detailed fine structures and the maximum flux are mainly due to the  diffraction efficiency of the grating. These differ between the calculated estimates of Fig.~\ref{Fig_transmission} and the real measurement. Namely the calculations predict higher flux around 60~eV and a less strong maximum. The key reasons for this discrepancy is likely to be a slight contamination of the grating surface, which shifts the best diffraction efficiency from the initially intended 40 to 60~eV to the actually observed value of 100 to 120~eV. 

\begin{figure}[!btp]
	\includegraphics[width = 0.42\textwidth]{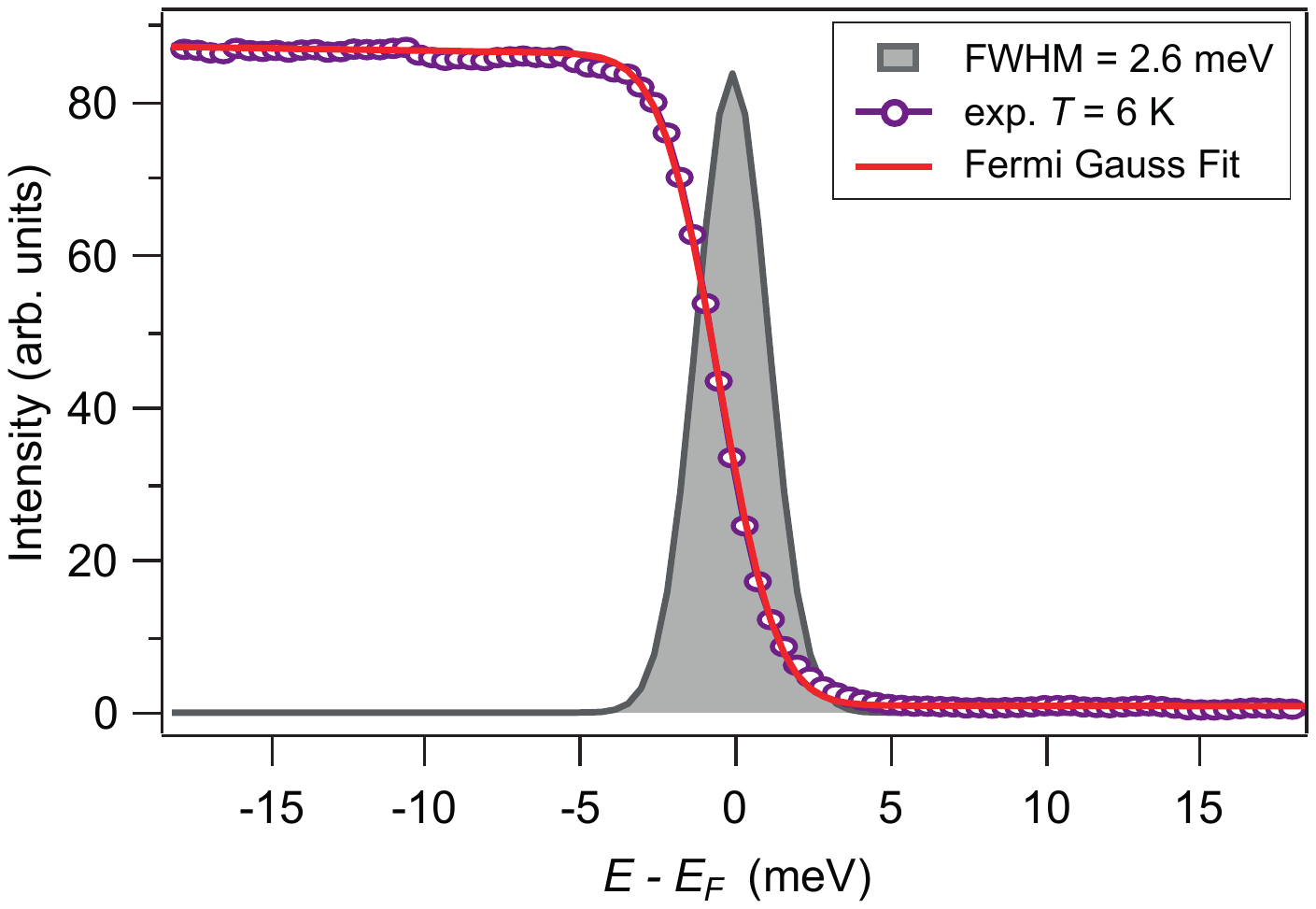}
	\caption{Measured photoemission spectrum from polycrystalline gold using a photon energy $h\nu = 20$~eV and exit slit $h=10$~$\mu$m, analysed as described in the text. A Gaussian function representing the combined experimental energy resolution of 2.6~meV is shown in grey.}
\label{fig_ana_energy_resol}
\end{figure}

With the beam thus delivered to the end station the energy resolution can now also be tested by photoemission spectroscopy. A typical spectrum from polycrystalline gold held at a temperature $T = 6$~K is shown in Fig.~\ref{fig_ana_energy_resol}. Beside the beam energy resolution $\Delta E$, which can be assumed to be unchanged from the exit slit to the sample, also the analyser resolution $\Delta E_{ana}$ as well as sample  and grounding imperfections ($\Delta E_{etc}$) contribute to the energy resolution. The spectrum is analysed by a fit to a Fermi-Dirac occupation function at temperature $T=6$~K convoluted with a Gaussian function representing the experimental energy resolution. This yields a combined resolution determined by a Gaussian with FWHM of $\Delta E_{comb} =  2.6$~meV.
 
\begin{figure}[!htbp]
	\includegraphics[width = .47\textwidth]{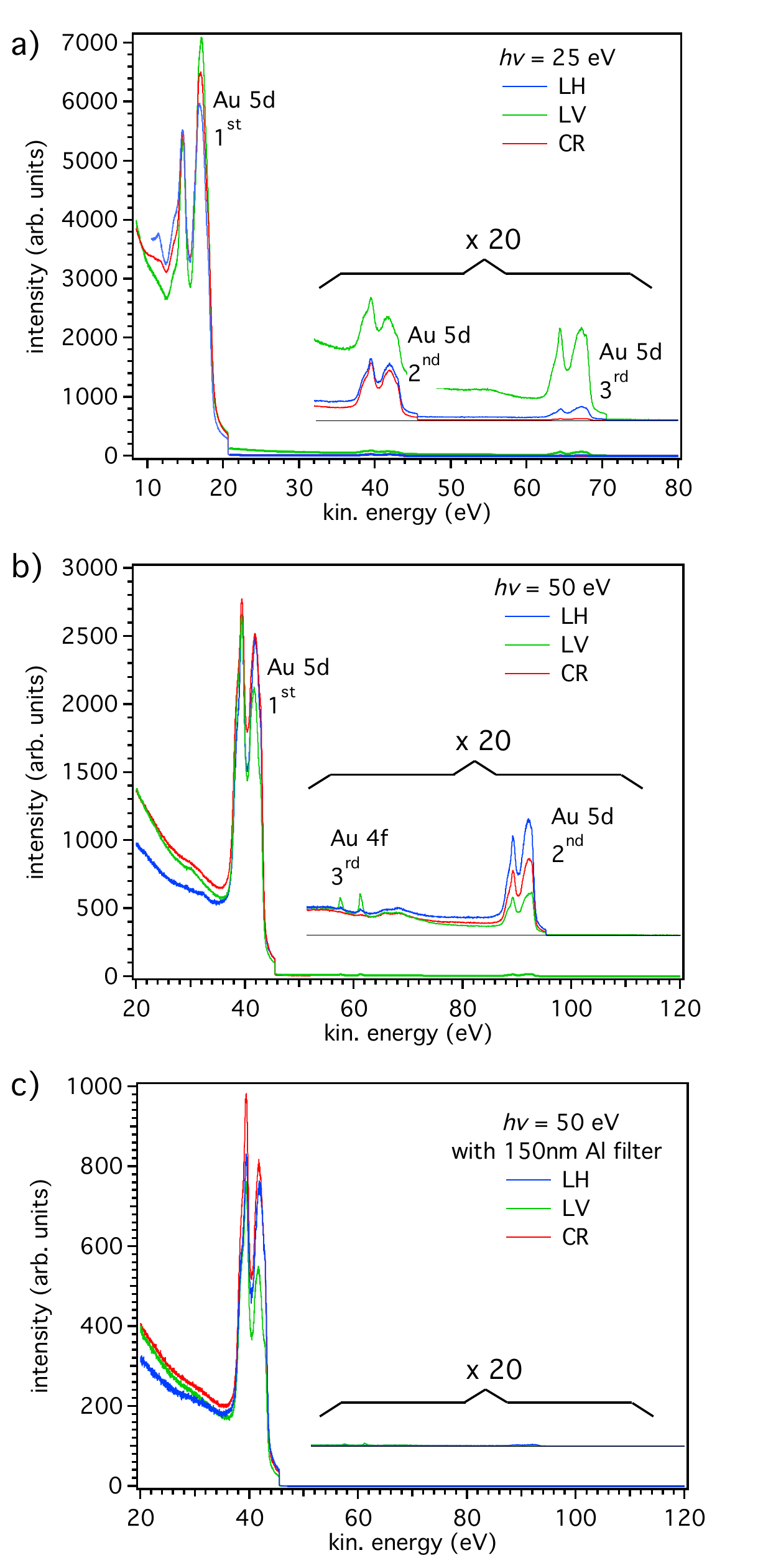}
	\caption{Measured spectrum of photoemission from polycrystalline gold for photon energy 25~eV (a) and 50~eV (b and c). The insert on the right side of each panel shows the same data multiplied by a factor 20. In (c) the beam passes through a 150~nm thick Al filter.  The intensity scales in (b) and (c) are normalised to the same absolute values. The labelling of peaks is described in the text.}
\label{Fig_harmonics}
\end{figure}

The higher harmonic contamination of the beam is now probed by the same photoemission experiments from polycrystalline gold. Data at photon energies of 25~eV and 50~eV at various polarisations are shown in Fig.~\ref{Fig_harmonics}. The peaks in the spectra are labelled according to the identified harmonic ($1^\mathrm{st}$, $2^\mathrm{nd}$ and $3^\mathrm{rd}$), and the Au subshell (4f and 5d).  A detailed interpretation of such spectra is complicated due to the fact that (a) the removed magnet blocks of the undulator play a different role depending on the phase motion of the magnet banks, (b) the different photoemission cross sections for different photon energies and (c) the energy-dependent transmission of the analyser. As a general trend the photoemission peaks corresponding to higher harmonics are at least a factor 100 smaller than the peaks from the fundamental, which have similar intensities for all polarisations. In the example of 25~eV the third harmonic is well visible in LH and LV polarisations and the ratio of $2^\mathrm{nd}$ to $3^\mathrm{rd}$ harmonics is similar, though LV shows more of both higher harmonics. At 50~eV, the $2^\mathrm{nd}$ harmonic is very weak in LV polarisation, while the $3^\mathrm{rd}$ harmonic is observed in LV but not in the other polarisations. A systematic rule of which harmonic to expect in which case is not derived. The background just above the Fermi step of the fundamental (just above 20~eV in (a) and 45~eV in (b) and (c)) is given by secondary electrons that relate to the higher harmonics peaks. This background is not suppressed to zero, but good signal-to-background is found in all cases. In the data shown in Fig.~\ref{Fig_harmonics} the worst case is in LV polarisation at $h\nu=25$~eV with a signal-to-background of 3.3. Finally the harmonics can be nearly completely suppressed by the insertion of the 150~nm thin Al filter, which absorbs photons above approx. 70~eV and thus acts as an excellent harmonics filter between 35 and 70~eV. This also reduces the fundamental, about 3 times at 50 eV (Fig.~\ref{Fig_harmonics} c), but may lead to higher quality data. 

\section{The end station instrument}
\label{sectEndstation}

\begin{figure*}[!tbhp]
	\includegraphics[width = .92\textwidth]{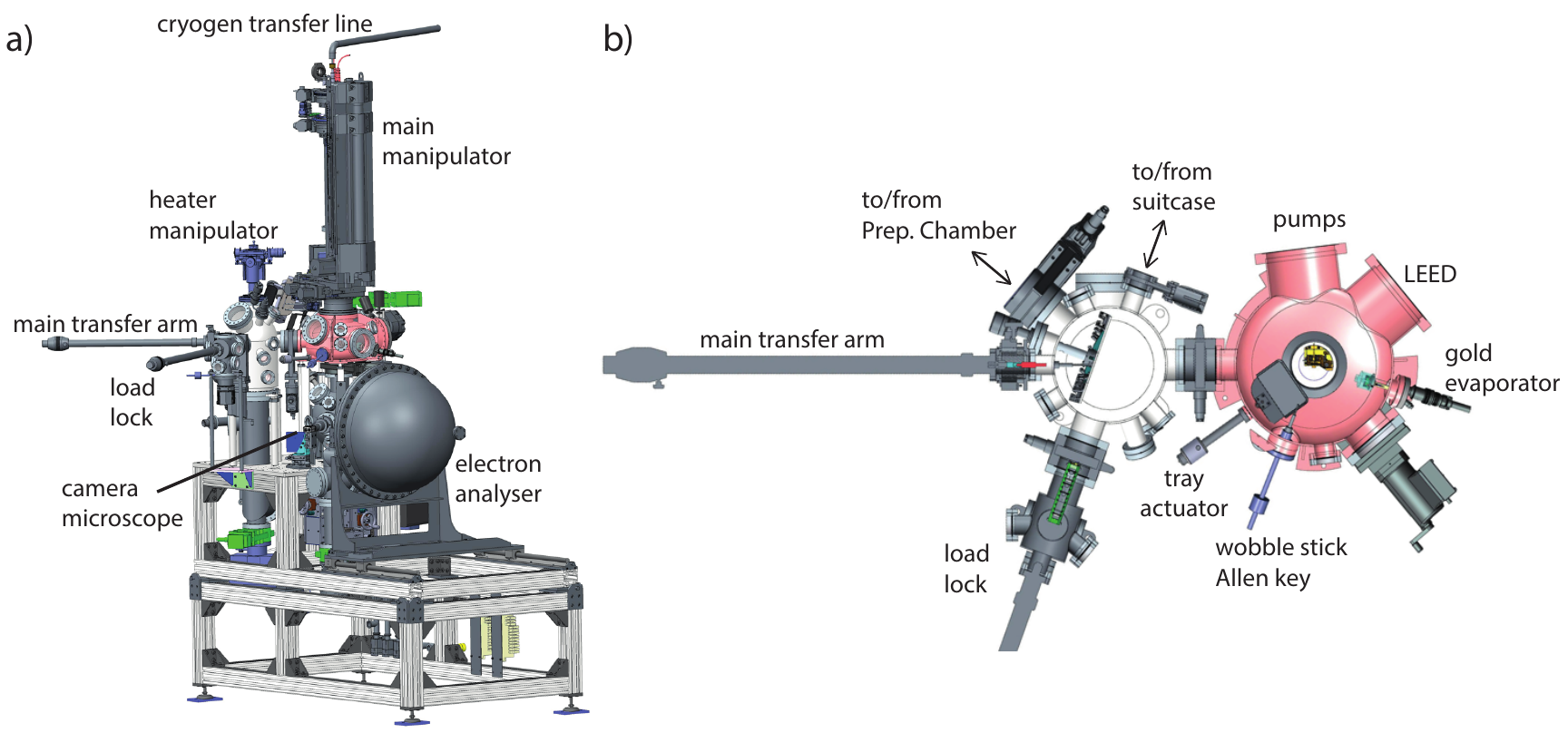}
	\caption{(a) Rendering of the HR-ARPES end station with the main vessels Lower Chamber (LC), Upper Chamber (UC), Interface chamber (IC) and load lock (LL) shown in grey, red, white and grey, respectively. (b) Cut through the transfer level of LL, IC and UC showing the multi-position load-lock recipient, the main transfer arm and the tray for capturing broken off cleavage posts in UC. In addition the sample storage disk in IC and additional ports on IC for transfers in and out of the Preparation Chamber and a vacuum suitcase are shown as well.}
\label{Fig_endstation}
\end{figure*}

The HR-ARPES end station consists of three ultrahigh vacuum (UHV) chambers, a fast entry load lock and a docking port for a UHV suitcase. Samples mounted on flag style holder plates are transferred between these chambers by magnetically coupled UHV transfer arms. The design of the station is shown in Fig.~\ref{Fig_endstation}. In 2014 the system was expanded by a surface science preparations and a molecular beam epitaxy system that can also be operated on their own in parallel with synchrotron radiation experiments~[\onlinecite{baker15}]. The end station needs to maintain very low residual gas levels of UHV as sample lifetime is usually limited by residual gas adsorption on the surface. The system is pumped by a combination of high compression ratio turbo molecular pumps, backed by scroll pumps, NEG pumps and small ion pumps. 

5 -10 samples at a time are inserted into the load lock and transferred to the interface chamber (IC) after a load lock pumpdown of 3 to 5 hours. The next transfer onto the cryogenic main manipulator in the upper chamber (UC) is performed after the vacuum levels in IC have recovered (5 minutes). The IC thus acts as a second stage of a two-stage load lock, keeping cryo-adsorption onto the manipulator low, which is manifest by low degassing on heating against the base temperature. With this procedure the vacuum levels in all chambers, except the load lock, are kept below $2\cdot10^{-10}$~mbar and sample lifetimes are highly competitive. The IC thus acts as a second stage of a two-stage load lock, keeping cryo-adsorption onto the manipulator low, which is manifest by low degassing on heating against the base temperature. Once loaded onto the main manipulator and cleaved (if appropriate) by removal of the attached top post, which is collected in a tray mounted in UC, the main manipulator is extended down into the lower chamber (LC) for measurements. The LC is made of $\mu$-metal for magnetic shielding and apart from the electron analyser and the beam line, both of which have additional pumping to reduce gas loads directed at the sample, the chamber only features window view ports for sample illumination and a camera microscope giving a constant view of the sample at an optical resolution of 5~$\mu$m.

\begin{figure*}[tbhp]
\includegraphics[width = 0.61\textwidth]{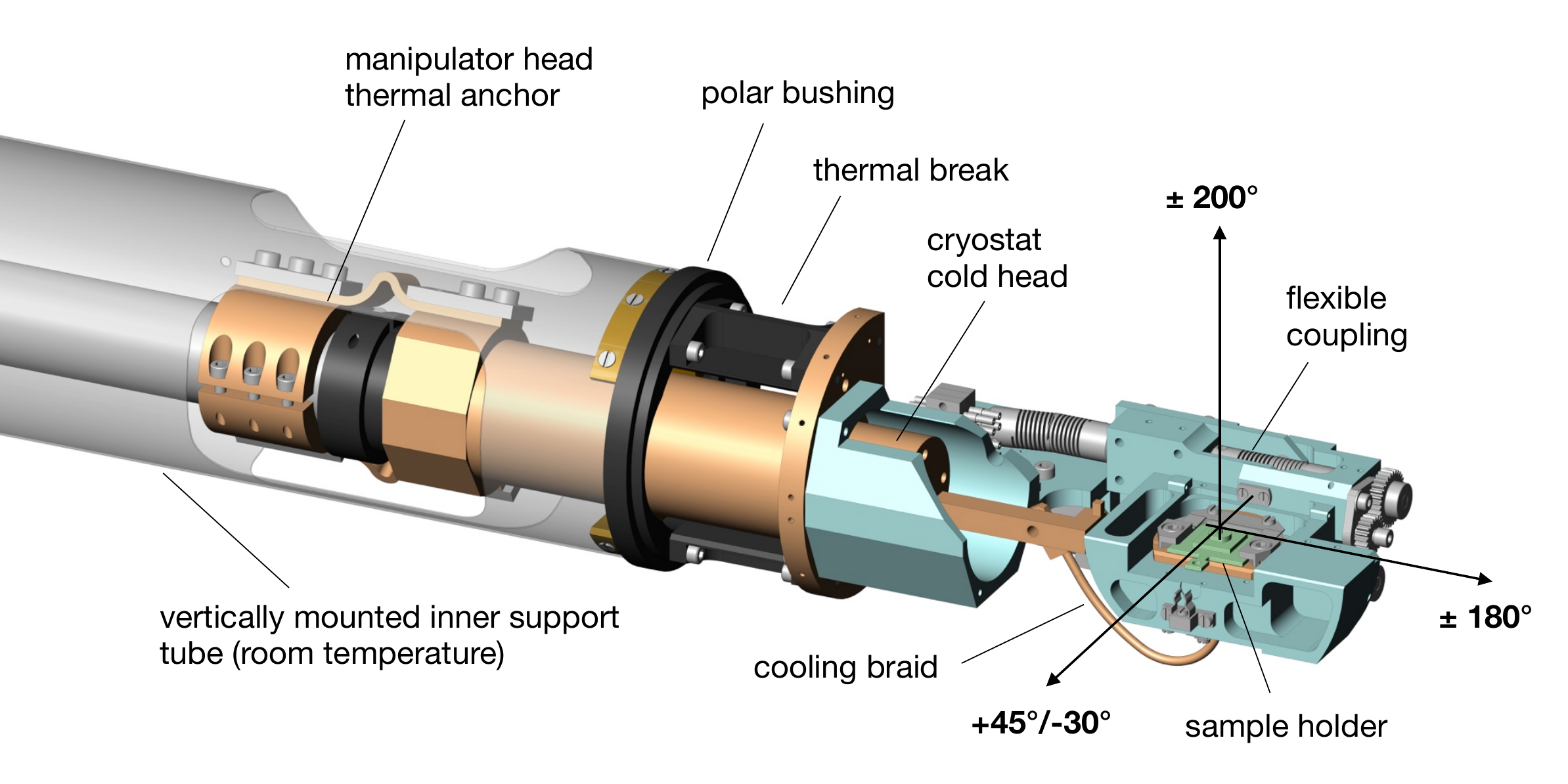} %
\caption{Rendering of the manipulator head. The three rotation axes with their respective angular ranges are indicated. The main support structure is machined in phosphor-bronze and shown in light blue. All thermal links (brown) are made in OFHC copper. Rotating parts are supported in polyimide (grade 2391 Tecasint) bushings. During ARPES measurements, the manipulator head is fully shielded with gold coated copper foil (not shown) to minimize radiative heat load on the sample and distortions of the electrostatic field between sample and analyzer entrance lens.}
\label{manipulator}
\end{figure*}

The analyser is a model R4000 from VG~Scienta company. The $\mu$-metal shielding of the analyser is joined to the LC by contact of the conical end piece into a matching surface of the LC port. The analyser features a standard lens, usually operated in angle-resolving mode thus allowing a parallel detection of electrons in a $\pm14\deg$ window.  A selection of both curved and straight entrance slits is available down to a slit opening of 100~$\mu$m. The detector is a combination of a multichannel plate amplifier with a fluorescence screen that is viewed by a CCD camera from outside of vacuum. The detector does not feature a mesh at its entrance, thus no imprint of the mesh structure is visible on the detector image and besides the traditional sweeping mode detection also a fixed mode of the detector is often used, which increases the speed of data acquisition.

Samples are oriented and aligned for ARPES measurements using a cryogenic sample goniometer with 6 degrees of freedom, cooled by a Janis ST-402 flow cryostat. A rendering of the goniometer head is shown in Fig.~\ref{manipulator}.
Three orthogonal translations are provided by a commercial bellow-sealed xyz-stage. Additionally, the sample can be rotated around three axes passing through the sample surface with a sphere of confusion of $\sim 200$~$\mu$m. The principal polar rotation around the long vertical axis is provided by a differentially pumped rotary seal with an additional bushing at the lower end of the support tube to center the axis in a second non-rotating support tube (not shown in Fig.~\ref{manipulator}) and reduce wobble of the sample position.
In order to minimize the heat load on the sample, the entire manipulator head is thermally decoupled from the room-temperature support tube and cooled by the exhaust gas of the cryostat. 
A fully independent tilt and azimuthal rotation is provided by a system of spur gears and worm shafts / helical wheels connected with backlash-free Inconel flexible couplings to rotary feedthroughs. The same flexible coupling is also used to transfer the shaft rotation from the feedthrough to the tilting azimuthal rotation axis.
The sample receptacle is cooled with a copper braid assembly and thermally insulated from the supporting mechanical system by a thin-walled polyimide cylinder (grade 2011 Tecasint, not visible in Fig.~\ref{manipulator}). A secondary cylindrical cold-shield surrounding the entire manipulator head is installed in the analysis chamber. This shield is cooled by an independent closed cycle cryostat (ColdEdge Stinger) and reaches a base temperature of 18~K, sufficient to condense all residual gases including H$_2$.

The mechanical performance of the goniometer has been characterized in UHV at low temperature. Using suitable backlash corrections, all rotation angles are reproducible to within {\color{black} $<0.05^{\circ}$ }, which is below the typical image distortion of modern electron spectrometers. Currently, the sample receptacle reaches a base temperature of 7~K with a flow of 0.38~l/h of liquid He. Experience with an identical design operated at the University of Geneva shows that a sample temperature of $T = 2.8$~K can be reached by pumping the cryostat and improving the thermal conductivity of the braid assembly. Sample temperatures up to 350~K are obtained using a heater near the cold-head of the cryostat. During temperature dependent measurements, thermal contraction of the head causes a translation of the sample position along the vertical axis of maximally $\sim300$~$\mu$m. In order to facilitate measurements of small samples, this drift can be corrected automatically with a software routine that uses the temperature of the manipulator support structure to control the vertical axis of the xyz-stage.

The HR-ARPES instrument is operated for approximately 30 user visits per year, lasting between 2 and 6 days and for inhouse research. In addition the beamline delivers beam to the nano-ARPES instrument and for beam optimisation studies. The high productivity of the instrument performance is best judged by the publication output~\cite{i05website}.

\section{Outlook}

The beamline and its HR-ARPES end station are fully commissioned and the performance of cryo-system, beam delivery resolution and flux and analyser resolution match each other. The photon flux is indeed so high that measurements with reduced flux are sometimes advisable, when the high intensity leads to detector saturation and energy resolution broadening due to space charge effects~\cite{zhou05}. This makes it attractive to further improve the energy resolution, which may be achieved by (a) further improved analyser resolution through the use of even smaller slits, (b) further improved reduction of electronic noise on the sample and (c) further improvement of the beam energy resolution through operation at higher $c$-values (thus stronger demagnification of the source) or installation of a grating with higher line density. 

The technique of ARPES may furthermore benefit from still improved detectors with flat response and higher linearity. For the typical high count rate operation of our instrument, the bottle-neck here is, however, the use of a multichannel-plate amplifier, which ages over time and thus develops deviations from flat response across the detector. The mechanical performance and reproducibility of the manipulator is so high that even smaller beam spot size, down to 5~$\mu$m would be an attractive development, thus enabling the measurement of smaller patches of sample within a cleaved surface. Furthermore the recent development of angle-sweeping lenses on the analyser would enable the measurement of relevant sections of momentum space without rotation or any movement of the sample and thus a faithful measurement from a single small spot on the sample surface.

With these developments, we consider ARPES a continuously productive technique over the next decade or two. Cleavable materials exist in abundance and new interest in low-dimensional systems will require the technique of ARPES for  detailed studies on novel samples. When cleavage is not possible, which is often the case in non-layered materials, then the in-situ growth of the material can still give a smooth, well-oriented surface and in this case the tuneability of momentum by the change of photon energy will continue to require synchrotron radiation light sources and beamlines of the kind described in this paper.

\begin{acknowledgements}
We wish to thank Diamond Light Source and its stakeholders for sponsoring this project under the leadership of G. Materlik, T. Rayment and R. Walker. A special thanks goes to C. Norris for paving the way to realisation of this installation and continuous support of the project throughout its lifetime with numerous discussions. The undulator design and construction was managed by J. Schouten and E. Rial kindly contributed further calculations to this manuscript. Expert advice was provided by K. Sawhney, J. Kay, N. Rees and M. Heron. The User Working Group (UWG) including A. Boothroyd, C. McConville, R. McGrath, C. Nicklin, and L. Patthey offered repeated reviews of the project and provided guidance and support. Fruitful discussions are acknowledged with R.~Reininger, R.~Follath, E.~Rotenberg, V.~Strocov, M.~Shi, D. H. Lu, P. D. C. King, Th. Hesjedal, A.~Taleb-Ibrahimi, U.~Flechsig, and T.~Schmidt. We are indebted for the excellent quality of components delivered from the main supplier companies Insync, Zeiss, SESO, Horiba Jobin-Yvon, Luxel, Mecaconcept, Bestec, FMB Oxford, Argon Services, Janis, ColdEdge, and VG~Scienta. One of us (MH) would like to acknowledge the kind hospitality of Hiroshima University during the time of writing this manuscript.

\end{acknowledgements}

\appendix
\section{Modelling of the gas absoption Fano line at large exit slit opening}

In Fig.~\ref{Fig_energy_resol}c), each spectrum is fitted using a convolution of a box with rounded edges, constructed from error functions erf, with the Fano profile:

\begin{widetext}
\begin{equation*}
i(\epsilon_t) = A \int^\infty_0\left[ \mathrm{erf}{}\left (\frac{(\epsilon_t-\epsilon)+\frac{D_ih}{2}}{\Delta_G} \right )  - 
\mathrm{erf}{}\left (\frac{(\epsilon_t-\epsilon) - \frac{D_ih}{2}}{\Delta_G} \right ) \right ] \sigma(\epsilon)d\epsilon; \; \;  \; \sigma(\epsilon) = \sigma_0+\sigma_F\frac{(q+\xi)^2}{1+\xi^2}; \; \;\; \xi =\frac{\epsilon-\epsilon_R}{\Delta_F/2}
\end{equation*}
\end{widetext}

Here $\epsilon_t$ is the tuneable energy of the monochromator, $D_i$ is the inverted linear dispersion at the exit slit, $h$ is the height of the exit slit. The parameter $\Delta_G$ represents the total instrument bandwidth (FWHM). This parameter defines the ultimate photon energy resolution that could be achieved. It is defined by at least the effects of photon source size, grating slope error and optical aberrations. The normalization amplitude $A$ equals to $F_0 nL$, where $F_0$ is the spectral flux density,  $n$ is the gas concentration, and $L$ is the length of the gas cell.
$\sigma(\epsilon)$ denotes the photoionization spectrum of He or Ne near the Fano resonance. Besides the offset $\sigma_0$, and amplitude $\sigma_F$ it is characterised by the resonant energy $\epsilon_R$, the resonance width $\Delta_F$ and the Fano asymmetry parameter $q$. These three parameters can be found in vast literature available on the subject (He [\onlinecite{domke96,schulz96}] and Ne [\onlinecite{klar92}]).
For the purpose of fitting we used tabulated values for  $\Delta_F$ and  $q$ of the Fano resonances. The tune energy $\epsilon_t$ and  exit slit height $h$ were measured arguments of the fit. The values $\sigma_0$, $\sigma_F$, $\Delta_G$, $\epsilon_R$, $A$ and $D_i$ were fitting parameters. According to the fitting model, the offset $\sigma_0$, and amplitude $\sigma_F$ of the Fano line are  parameters of the gas and so should not vary with the exit slit height. Indeed, only slight variations of these two parameters were observed. The value of the inverse linear dispersion found by the fit is 86 meV/mm, which corresponds well to the theoretical estimate~\cite{petersen95}.

\bibliography{beamline}

\end{document}